\documentclass[10pt, conference]{IEEEtran}
\IEEEoverridecommandlockouts
\usepackage[noadjust]{cite}
\usepackage{amsmath, amssymb, amsfonts}
\usepackage{algorithmic}
\usepackage{graphicx}
\usepackage{textcomp}
\usepackage{xcolor}

\def\BibTeX{{\rm B\kern-.05em{\sc i\kern-.025em b}\kern-.08em
    T\kern-.1667em\lower.7ex\hbox{E}\kern-.125emX}}

\begin{document}

\title{Frequency Modulation for Task-Oriented Communications and Multiple Access
\thanks{This work is part of the project IRENE (PID2020-115323RB-C31), funded by MCIN/AEI/10.13039/501100011033 and supported by the Catalan government through the project SGR-Cat 2021-01207.\\
An extended version of this paper was presented at IEEE GLOBECOM 2023.}
}

\author{\IEEEauthorblockN{Marc Martinez-Gost\IEEEauthorrefmark{1}\IEEEauthorrefmark{2}, Ana Pérez-Neira\IEEEauthorrefmark{1}\IEEEauthorrefmark{2}\IEEEauthorrefmark{3}, Miguel Ángel Lagunas\IEEEauthorrefmark{2}}
\IEEEauthorblockA{
\IEEEauthorrefmark{1}Centre Tecnològic de Telecomunicacions de Catalunya, Spain\\
\IEEEauthorrefmark{2}Dept. of Signal Theory and Communications, Universitat Politècnica de Catalunya, Spain\\
\IEEEauthorrefmark{3}ICREA Acadèmia, Spain\\
\{mmartinez, aperez, malagunas\}@cttc.es
}}

\maketitle
\begin{abstract}
In the context of task-oriented communications we advocate the development of waveforms for Federated Edge Learning (FEEL). Over-the-air computing (AirComp) has emerged as a communication scheme that allows to compute a function out of distributed data and can be applied to FEEL. However, the design of modulations for AirComp is still in its infancy and most of the literature ignores this topic. In this work we employ frequency modulation (FM) and type based multiple access (TMBA) for FEEL and demonstrate its advantages with respect to the state of the art in terms of convergence and peak-to-average power ratio (PAPR).
\end{abstract}

\section{Introduction}

In a wireless sensor network (WSN) nodes gather information from their surroundings with the ultimate goal of data fusion. This demands the development of task-oriented communications, this is,
exploiting the semantic relationships and folding the communication goals into the design of task-level transmission strategies \cite{shi23_task}.
In the case of distributed estimation tasks, the communication must be designed to produce the best estimate (e.g., minimum mean squared error).
The development of task-oriented communications is motivated by the increasing interest in  distributed learning schemes, specifically in Federated Edge Learning (FEEL). In FEEL a set of $K$ devices train a shared model in a collaborative fashion \cite{gaf22}. At each communication round, each device trains a local model on its own data and then shares the model parameters with the server. The latter generates a global model merging (e.g., averaging) the received parameters, which is then distributed back to the devices for further updates until convergence.


A primary obstacle in Federated Edge Learning (FEEL) involves managing limited resources, such as power, at the edge devices. Transmitting large data volumes across the network becomes challenging under these constraints, resulting in excessive delays and compromising the overall efficiency of the learning process.
Thus, developing task-oriented communications for FEEL suggests designing waveforms and multiple access schemes that allow to transmit the parameters for the purpose of training a distributed model, not pure communication (i.e., minimum error probability).
Over-the-air computing (AirComp) is a task-oriented transmission that has gained relevance in the context of FEEL. When all users transmit synchronously in time and frequency, the multiple access channel (MAC) yields a superimposed signal that is proportional to the mean, which is the amount of interest at the receiver \cite{Nazer2007}.

In \cite{Zhu19} the authors propose an analog scheme where each device modulates the information in the amplitude of the symbol and enables simultaneous transmission. However, these modulations (e.g., double-sideband modulation, DSB) may pose challenges, particularly at the transmitter side: adjusting the power of the carrier due to variations in the modulated information may push the amplifier outside its operational range. Also, edge devices may not be able to provide the transmission power that linear analog modulations require.
To address these issues, \cite{zhu20, sahin21} propose digital schemes. These approaches utilize binary frequency shift keying (BFSK) to transmit only the sign of gradients, with the server resolving the aggregated gradient through majority voting. While these schemes show empirical convergence at high signal-to-noise ratio (SNR), low SNR scenarios may require multiple communication rounds. Furthermore, these schemes rely on gradient-averaging, but it is cumbersome to incorporate model-averaging.

\section{FM for distributed wireless computing}
To address the previous issues, we propose a frequency modulation (FM) aggregation for FEEL.
Specifically, we generalize the work in \cite{zhu20, sahin21} to incorporate the magnitude of either the parameter or the gradient and speed up the learning process.

The $q$-th parameter of the $k$-th local model, $\mathbf{w}_k[q]$, is quantized as
\begin{equation}
    m_{k,q}=g\left(\mathcal{Q}\left(\mathbf{w}_k[q]\right)\right)\in [0,\dots,N-1],
    \label{eq: quantizer}
\end{equation}
where $\mathcal{Q}$ is an $N$-level uniform quantizer and function $g$ maps the quantized parameter to $[0,N-1]$. Each model parameter is then modulated with an $M$-ary Frequency Shift Keying (MFSK) as
\begin{equation}
z_{k,q}[n] = A_c\sqrt{\frac{2}{N}} \cos\left(\frac{\pi(2m_{k,q}+1)}{2N}n\right),
\label{eq:dct_tbma}
\end{equation}
for $n=0,\dots,N-1$. The waveform in \eqref{eq:dct_tbma} corresponds to a single tone whose frequency is assigned depending on the parameter magnitude. This is the core of type-based multiple access (TBMA), a semantic-aware multiple access scheme where resources are assigned to orthogonal parameters, not users. TBMA lies between pure AirComp and orthogonal multiple access \cite{Mergen2006}. Specifically, AirComp occurs when two users modulate the same information ($m_{k,q}=m_{k',q}, k\neq k'$), as they generate the same waveform and the channel provides a constructive interference.
Furthermore, TBMA assigns communication resources in a fully distributed fashion.

The signal at the input of the receiver is
\begin{equation}
    y_q[n] = \sum_{k=1}^K z_{k,q}[n] + w[n],
    \label{eq:tbma}
\end{equation}
where $w$ corresponds to additive white Gaussian noise (AWGN) samples. Computing the matched filter of \eqref{eq:tbma} at the $m$-th frequencies, corresponds to a noisy version of the number of users that transmitted that tone. Thus, \eqref{eq:tbma} in the frequency domain results in a noisy version of the histogram (or type) for parameter $q$, over which the mean can be computed.

While TBMA is not limited to MFSK (it only requires $N$ orthogonal waveforms), we choose MFSK because FM is more robust to noise and more energy efficient due to its constant envelope. Besides, linear analog modulations cannot guarantee a certain SNR level, because it requires prior knowledge of all the transmitted data, which happens to be distributed. On the other hand, since FM waveforms are designed to have unit power, the SNR can be achieved effortlessly.
From a computing perspective, TBMA generalizes AirComp. Specifically, any calculation that is a function of the type can be computed. This includes, for instance, alternative means (harmonic, geometric...) and statistics (maximum, minimum, variance...).



\section{Results}
\label{sec:experiments}
We consider the deployment of a FEEL system with $K=50$ devices in an AWGN channel and standard parameter averaging is computed. As the proposed scheme requires coherent demodulation, we assume the fading is already compensated.
The task is image classification using the MNIST dataset and the learning model consists in a standard convolutional neural network (CNN). The performance is measured by the average test accuracy (i.e., percentage of test samples correctly classified) across all devices.

\begin{figure}[t]
\centering
\includegraphics[width=\columnwidth]{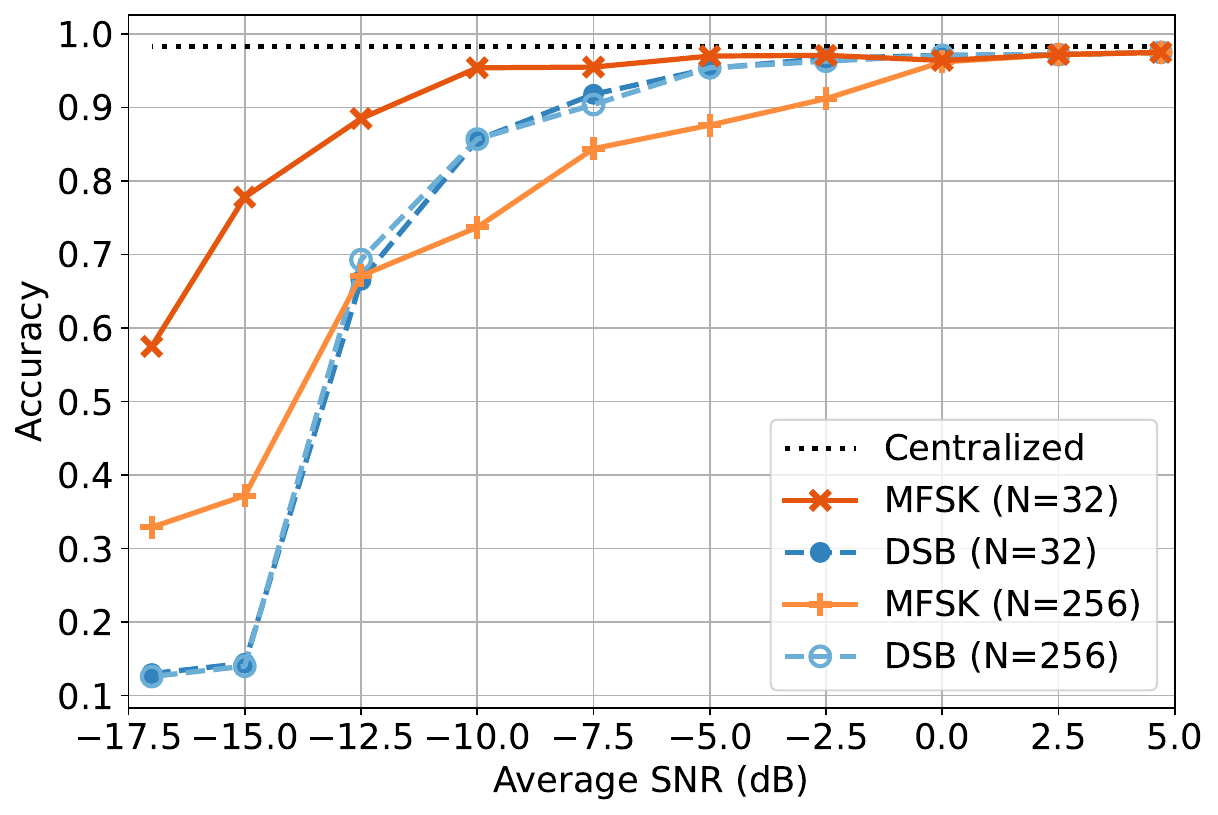}
\caption{Accuracy of the FEEL system for the digital frequency modulation (MFSK) and the linear analog modulation (DSB) for $N=\{32,256\}$, along with the upper bound of a centralized trained model.}
\label{fig:sim_accuracy_SNR}
\vspace{-0 pt}
\end{figure}

Fig. \ref{fig:sim_accuracy_SNR} shows the accuracy achieved by TBMA implemented with MFSK and for linear analog AirComp with DSB modulation for $N=\{32,256\}$. The performance of a centralized model with no FEEL architecture and no communication scheme is presented as well as an upper bound on the model performance ($98\%$).
As expected, MFSK requires $K>N$ to recover an accurate approximation of the histogram. In this case, the performance of the frequency modulation remains almost intact up to $-10$ dB, which does not happen for the DSB scheme. Below $-10$ dB the noise corrupts the mean estimation, which traps the algorithm in a local minima or prevents it from converging.
Regarding the power efficiency, the peak-to-average power ratio (PAPR) of MFSK is 0 dB (i.e., constant envelope), while DSB achieves 14 dB. Moreover, a high PAPR signal may be more susceptible to distortion and interference, which can affect the performance of the system.

\section{Future work}
We envision that TBMA is a powerful task-oriented communication scheme for many distributed estimation problems. 
From a signal processing perspective, we will analyze the effect of different averaging techniques and propose TBMA for robust learning. From the communication side, we will integrate fading channels and deepen in the relationship between transmission bandwidth and convergence speed, which relates to the schemes proposed in \cite{zhu20, sahin21}.

\bibliographystyle{IEEEbib}
\bibliography{refs}

\end{document}